\documentclass[aps,pra,twocolumn,halfspacing,10pt,nofootinbib]{revtex4-1}
\usepackage{epsfig}
\usepackage{dcolumn}
\usepackage{natbib}
\usepackage[colorlinks]{hyperref}
\usepackage[utf8]{inputenc}
\usepackage[english]{babel}
\usepackage{graphicx}
\usepackage{color}
\usepackage{bm}
\usepackage{amssymb}
\usepackage[intlimits]{amsmath}
\usepackage{amsmath}
\usepackage[para,online,flushleft]{threeparttable}

\begin{document}
\title{Sensitivity of isotope shift to  distribution of nuclear charge density}
\author{V. V. Flambaum$^{1,2}$}
\author{V.A. Dzuba$^1$}
\affiliation{$^1$School of Physics, University of New South Wales,
Sydney 2052, Australia}
\affiliation{$^2$Helmholtz Institute Mainz, Johannes Gutenberg University, 55099 Mainz, Germany}

\date{\today}

\begin{abstract}
It is usually assumed that  the field isotope shift (FIS) is completely determined by the change of the averaged squared values of the nuclear charge radius  $\langle r^2\rangle$.
Relativistic corrections modify the expression for FIS, which is actually described by the change of $\langle r^{2 \gamma}\rangle$, where $\gamma=\sqrt{1 - Z^2 \alpha^2}$.
In the present paper we consider  corrections to FIS which are due to the nuclear deformation and due to  the predicted reduced charge density in the middle of the superheavy nuclei produced by a very strong proton repulsion  (hole in the nuclear centre). Specifically, we investigate effects which can not be completely reduced to the change of $\langle r^2 \rangle$ or $\langle r^{2 \gamma}\rangle$. 
\end{abstract}

\maketitle

\section{Introduction}

Isotope shift (IS) phenomena in heavy atoms are an important way of probing various scenarios in nuclear physics and can aid the search for new physics beyond the Standard Model. Nuclear theory predicts the existence of long-lived isotopes for elements with $Z\geq 104$ (see e.g. \cite{oganessian_heavy_2004,hamilton_search_2013}), in particular isotopes with a magic neutron number $N=184$. However, producing these neutron-rich isotopes in laboratories by colliding lighter atoms is currently impossible. The Coulomb repulsion for nuclei grows as $Z^2$; in order to compensate for this with the attractive strong force, the neutron number $N$ must 
grow faster than $Z$. Consequently, an isotope from the island of stability with $N=184$ cannot be produced from the collision of a pair of lighter isotopes with smaller $N/Z$ ratios.

In contrast to laboratories, various astrophysical events such as supernovae explosions, neutron stars and neutron star - black hole/neutron star mergers generate high neutron fluxes and may create environments favourable for the production of neutron-rich heavy elements.
For example, a new mechanism of such a kind due to the capture of the neutron star material by a primordial black hole has been suggested in 
\cite{fuller_primordial_2017}. Furthermore, neutron star - neutron star mergers are predicted to generate optimal environments for the production of heavy atoms \cite{goriely_r-process_2011,frebel_formation_2018}. 

As a consequence, astrophysical data may be the best place to observe super-heavy meta-stable elements. It is possible that optical lines of elements up to $Z=99$ have already been identified in the spectra of Przybylski's star \cite{gopka_identification_2008}. These elements include heavy, short-lived isotopes which may be products of the decay of long-lifetime nuclei near the island of stability \cite{dzuba_isotope_2017}. 

IS calculations for superheavy elements (SHE) can help trace the hypothetical island of stability in existing astrophysical data. It may be possible to predict a spectral line of a neutron-rich isotope $\nu '$ based on the experimental spectrum of a neutron-poor isotope $\nu$ and calculations of IS $\delta \nu$ as $\nu ' = \nu + \delta \nu$. The results can then be used to search for the long-lifetime neutron-rich elements in complicated astrophysical spectra such as that of Przybylski's star.

Spectroscopic measurements of IS may also be relevant to the search for strange-matter. Strange nuclei consist of up, down and strange quarks (see \cite{witten_cosmic_1984} and references therein).
A strange-matter nuclei of charge $Z$ would have a very different radius in comparison to any regular isotope. Calculations of IS can be used to predict the effects of this change in radius on atomic spectra.

Calculations of IS  allows one to estimate the King-plot nonlinearity of a given element. New long-range forces such as Yukawa-type interactions between electrons and nucleus can lead to nonlinearities in a King plot for a series of isotopes \cite{berengut_probing_2017}. 
It is useful to understand other possible sources of nonlinearities in the IS in order to 
constrain new physics beyond the Standard Model.

It should be noted that relativistic corrections produce an important difference in the dependence of the field shift on the nuclear radius $r$. The traditional expression for field shift is known as $F_i \delta\left<r^2\right>$ where $F_{i}$ is an electronic structure factor and $\delta\left<r^2\right>$ is a nuclear parameter. It is usually assumed that electron factor
$F_i$ is the same for all isotopes. In fact, relativistic effects break this independence and if the independence on isotopes
is to be kept the field shift should be written as ${\tilde F}_i \delta \left<r^{2 \gamma}\right>$, where 
$\gamma=\sqrt{1 - Z^2 \alpha^2}$,  $\alpha$ is the fine structure constant. The electronic factor ${\tilde F}_i$ is to be calculated. Analytical estimate of ${\tilde F}_i$  has been done in  Ref.~\cite{FlambaumGeddesViatkina} (see also
\cite{racah_isotopic_1932,rosenthal_isotope_1932,shabaev_finite_1993}), relativistic many body calculations for $Z=102-109$ have been done in Refs. \cite{No,Db,Sg-Mt}.
The traditional formula for the field shift $F_i \delta \left<r^2\right>$ still can be used for neighbouring isotopes where
change in $F_i$ is small and can be neglected. The formula is useful for finding the change in nuclear root mean square (RMS) radius from the IS
measurements. 

Due to the relativistic effects in heavy atoms, the field shift of the $p_{1/2}$ orbital is comparable to that of the $s_{1/2}$: the ratio is $\sim (1- \gamma)/(1+\gamma)$~\cite{FlambaumGeddesViatkina}. The $ Z\alpha$ expansion gives the ratio $\sim Z^2 \alpha^2/4$ but for $Z$=137, $\gamma \approx 0$ and for the superheavy elements the ratio tends to 1. For $j>1/2$ the direct mean-field single-particle field shift is small. However, the mean-field rearrangement effect (the correction to the atomic potential $\delta V$ due to the perturbation of the $s$ and $p_{1/2}$ orbitals by the field-shift operator) 
produces the same dependence of field shift on nuclear radius for all orbitals:
${\tilde F}_i \delta \left<r^{2 \gamma}\right>$.


The difference between the non-relativistic  $\langle r^2\rangle$ and relativistic  $\langle r^{2 \gamma}\rangle$  expressions may be explained by the different dependence of the non-relativistic and relativistic wave functions near the origin.   Another relativistic effect is due the variation of the electron density $\rho_e$ inside the nucleus which  for the $s$ and $p_{1/2}$ orbitals is approximately presented by the following formula \cite{FlambaumGeddesViatkina}:
\begin{equation}
\label{rhoIn}
\rho_e(r) \approx  \rho_e(0) \left(1-\frac{Z^2 \alpha^2 }{2}  \left(\frac{r}{c}\right)^{2}\right)\,
\end{equation}
where $c$ is the nuclear radius. 
The $r$-dependent term gives us an additional sensitivity of IS to the nuclear charge distribution beyond the change of $\langle r^2 \rangle$. 

In this work we study the effect of the change in nuclear charge distribution on the field isotope shift. We consider
four types of charge distribution variation: (a) a hole in the origin, where nuclear density is small in the origin and increases to the periphery; (b) nuclear quadrupole deformation; (c) change of the skin thickness; and (d) change in nuclear RMS radius.
The questions we try to answer include (a) can isotope measurements be used to study nuclear structure beyond the change of nuclear RMS radius; (b) what is the best way of using isotope shift calculations to predict the spectra of neutron-rich SHE with the aim to reach the hypothetical island of stability; (c) can nuclear deformation lead to non-linearity of King plot.

We choose the E120$^+$ ion for numerical analysis. It is sufficiently heavy for the relativistic effects to be pronounced.
On the other hand the ion has relatively simple electron structure (one external electron above closed shells) so that all 
important points can be illustrated without getting into a trouble of complicated many-body calculations. 
We use the results of nuclear calculations~\cite{Afanasiev} to get the parameters of nuclear deformation and nuclear RMS radius.
We consider only even isotopes because nuclear calculations for them are more reliable. The work~\cite{Afanasiev} considers a range of nuclear models which favour spherical nuclear shape at $Z=120$ and $N=172$. We use this spherical nucleus as starting point in our study.

\section{Calculations}

We use an approach similar to one in Ref.~\cite{E120,E120a}. Electron potential $V$ for valence orbitals is found by solving 
relativistic Hartree-Fock (RHF) equations for a closed-shell core
\begin{equation}
(\hat H^{\rm HF} -\epsilon_c) \psi_c =0,
\label{e:HF}
\end{equation}
where $c$ numerates states in the core from $1s$ to $7p_{1/2}$ and $7p_{3/2}$. States of valence electron (Brueckner
orbitals) are obtained by solving the RHF-like equations for the valence orbitals
\begin{equation}
(\hat H^{\rm HF} +\lambda \Sigma^{(2)}-\epsilon_v) \psi^{\rm Br}_v =0.
\label{e:Br}
\end{equation}
Here $\Sigma$ is the correlation potential responsible for core-valence correlations \cite{Sigma}, index "2" indicates second order of the many-body perturbation theory. 
$\Sigma$ is defined in such a way that the correlation correction to the energy $\epsilon_v$ is given by 
$\delta \epsilon_v = \langle \psi_v|\Sigma |\psi_v\rangle$ (see, e.g. \cite{Sigma} for details).
We calculate $\Sigma$  {\em ab initio}, limiting ourselves to the lowest order of the perturbation theory. 
$\lambda$ is a scaling parameter introduced to simulate
the effect of higher-order correlations. Its value ($\lambda=0.75$) is chosen to fit the result of all-oder calculations
of Ref.~\cite{E120,E120a}. 

IS is calculated using the so-called random phase approximation (RPA, see e.g. \cite{Sigma}) which can be described as linear response 
of self-consistent atomic field to a small perturbation. In our case the perturbation is the change in nuclear potential
$\delta V_N$ due to change in nuclear charge distribution.
The RPA equations are first solved for the core 
\begin{equation}
(\hat H^{\rm HF} -\epsilon_c) \delta \psi_c =-(\delta V_N +\delta V_{core}),
\label{e:RPA}
\end{equation}
where $\delta \psi_c$ is the correction to the core orbitals due to the effect of $\delta V_N$, $\delta V_{core}$ is the
correction to the electron potential of core electrons due to the changes in all core orbitals. IS for states of 
a valence electron is found as $\langle \psi_v^{\rm Br}| \delta V_N + \delta V_{core} | \psi_v^{\rm Br} \rangle$.

\begin{figure}
\epsfig{figure=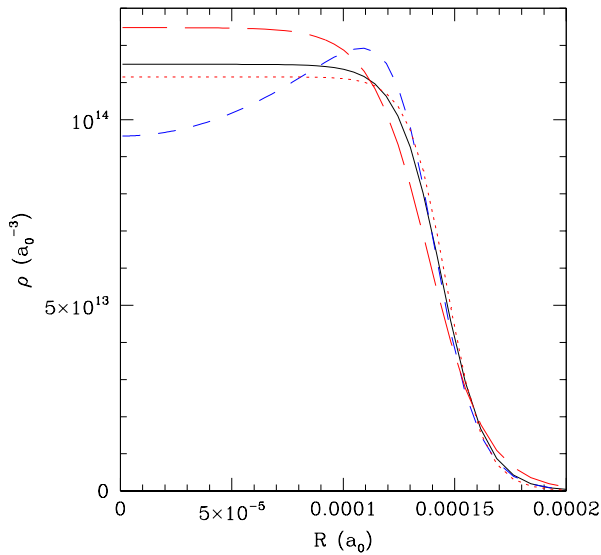,scale=0.4}
\caption{Variations of nuclear density. Solid (black) line - Fermi distribution (\ref{e:Fermi});
dashed (blue) line - modified distribution with a hole in the origin, formula (\ref{e:hole}) with $k=0.5$;
dotted (red) line - Fermi distribution with reduced skin thickness (parameter $t$ in (\ref{e:Fermi})) by 14.5 \% to
simulate the effect of the hole; long dashed (red) line  - Fermi distribution with inclreased skin thickness by 30.5 \% to
simulate the effect of quadrupole deformation, formula (\ref{e:beta}) with $\beta=-0.4$.}
\label{f:dro}
\end{figure}

We use Fermi nuclear charge distribution (solid line on Fig.~\ref{f:dro})
\begin{equation}
\rho(r)_f = \frac{\rho_0}{1+\exp{4\ln 3 (r-c)/t}},
\label{e:Fermi}
\end{equation}
where $c$ is nuclear radius, $t$ is skin thickness, and $\rho_0$ is normalisation constant, $\int \rho(r)_f dV =Z$.
Nuclear charge distribution with a hole in the origin is given by (dashed line on Fig.~\ref{f:dro})
\begin{equation}
\rho(r)_h = \rho(r)_f \left(1+k\left(\frac{r}{c}\right)^2\right).
\label{e:hole}
\end{equation}
The normalisation constant $\rho_0$ is adjusted to keep correct normalisation.
Nuclear quadrupole deformation is considered by replacing constant nuclear radius $c$ in (\ref{e:Fermi}) by varying
parameter $c(\theta)$
\begin{equation}
c(\theta) = c\left(1+ \beta Y_{20}(\theta)\right).
\label{e:beta}
\end{equation}
and calculating spherical average by integrating over $\theta$. It is known that this is approximately equivalent to increase in skin
thickness \cite{Clark,Heisenberg}
\begin{equation}
t^2 \approx t^2_0 + \left(4\ln3\right)^2\left(3/4\pi^3\right)c^2\beta^2.
\label{e:beta1}
\end{equation}
We also consider the change of nuclear radius. We use the $^{292}E120$ isotope as a reference one and we take 
nuclear parameters from nuclear calculations \cite{Afanasiev}.

\begin{figure}
\epsfig{figure=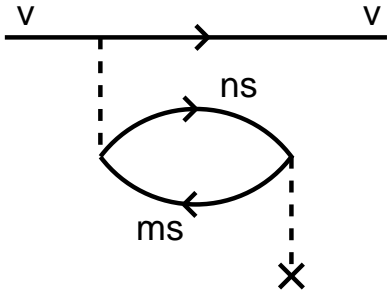,scale=1.0}
\caption{Dominating contribution to the isotope shift of the single-electron states $v$ with total angular momentum $j>1/2$ ($p_{3/2}, d_{3/2}, d_{5/2}$, etc). Cross stands for $\delta V_N$, change of nuclear potential due to change in nuclear charge distribution.}
\label{f:rpa}
\end{figure}

\section{Results} 

\begin{table}
\caption{\label{t:E120}Nuclear parameters for the  range of even isotopes from $^{292}$E120 to $^{306}$E120 isotopes taken from~\cite{Afanasiev}.}
\begin{ruledtabular}
\begin{tabular}{ldd}
\multicolumn{1}{c}{$A$}&
\multicolumn{1}{c}{$\beta$}&
\multicolumn{1}{c}{$\sqrt{\langle r^2 \rangle}$ (fm)}\\
\hline
292 & 0.0 &  6.220 \\
294 & -0.174 & 6.264  \\
296 & -0.205 & 6.294  \\
298 & -0.218 & 6.330  \\
300 & -0.221 & 6.358  \\
302 & -0.261 & 6.297  \\
304 & -0.290 & 6.484  \\
306 & -0.376 & 6.503  \\
\end{tabular}
\end{ruledtabular}
\end{table}

Table~\ref{t:E120} lists isotopes of SHE E120 used in this study with nuclear parameters taken from~\cite{Afanasiev}.
The results are presented in Tables~\ref{t:IS} and \ref{t:w} and Fig.~\ref{f:dv0}.
In all cases the IS for $s$ and $p_{1/2}$ states is dominated by the 
$\langle \phi_a^{\rm Br}| \delta V_N | \phi_a^{\rm Br} \rangle$ term (see Table~\ref{t:IS}); IS for states with $j>1/2$ is dominated by the
core polarisation (CP) term $\langle \phi_a^{\rm Br}| \delta V_{core} | \phi_a^{\rm Br} \rangle$. The largest contributions 
to the CP comes from the core $s$ states as shown on Fig.~\ref{f:rpa}. Therefore, the effect of change in nuclear 
charge distribution is very similar for all states except the $p_{1/2}$ states. 

\subsection{Hole in nuclear charge distribution and change of the nuclear skin thickness}

\begin{table*}
\caption{\label{t:IS}Isotope shift for specific states of E120$^+$ ( in $10^{-3}$~cm$^{-1}$) due to change in 
nuclear charge distribution. Reference IS is the IS between $^{292}$E120 and $^{294}$E120 calculated ($\langle \psi_a^{\rm Br}| \delta V_N + \delta V_{core} | \psi_a^{\rm Br} \rangle$) with the nuclear parameters from Table~\ref{t:E120}. "Hole" is the shift due to the difference between pure Fermi distribution (\ref{e:Fermi}) 
and the distribution with the hole in the origin, formula (\ref{e:hole}) with $k=0.5$. The same IS is produced by reducing
the skin thickness $t$ in (\ref{e:Fermi}) by 14.5\%. "Deformation" is the shift due to quadrupole deformation, formula 
(\ref{e:beta}) with $\beta=-0.4$. The same IS is produced by increasing the skin thickness $t$ in (\ref{e:Fermi}) by 30.5\%.
Note that while changing the hole parameter $k$ or the deformation parameter $\beta$ we are also changing nuclear radius parameter $c$ to keep the rms radius unchanged. 
"Change of $\sqrt{\langle r^2 \rangle}$" is the IS due to change of nuclear RMS radius in pure Fermi distribution (\ref{e:Fermi}) from 6.220~fm to
6.211fm. "Br" stands for IS given by $\langle \psi_a^{\rm Br}| \delta V_N | \psi_a^{\rm Br} \rangle$; "Br+CP" includes core polarization, $\langle \psi_a^{\rm Br}| \delta V_N + \delta V_{core} | \psi_a^{\rm Br} \rangle$. 
Note that corresponding matrix elements may be interpreted as isotope shift corrections to the ionisation potential for an electron on a given orbital.  
}
\begin{ruledtabular}
\begin{tabular}{l rrrrrrr}
\multicolumn{1}{c}{State}&
\multicolumn{1}{c}{Reference}&
\multicolumn{2}{c}{Hole}&
\multicolumn{2}{c}{Deformation}&
\multicolumn{2}{c}{Change of $\sqrt{\langle r^2 \rangle}$}\\
&\multicolumn{1}{c}{IS}
&\multicolumn{1}{c}{Br}
&\multicolumn{1}{c}{Br+CP}
&\multicolumn{1}{c}{Br}
&\multicolumn{1}{c}{Br+CP}
&\multicolumn{1}{c}{Br}
&\multicolumn{1}{c}{Br+CP} \\
\hline
$8s$       &    10134 &    743 &  813 &          -1986 & -2172 &          -1988 & -2172 \\
$9s$       &      2377  &   182 &  191 &           -486 &  -510 &           -486 &  -510 \\
$8p_{1/2}$ &   1705 &    130 &  131 &           -347 &  -351 &           -359 &  -365 \\
$8p_{3/2}$ &   -485 & $\sim 10^{-2}$ &  -38 & $\sim 10^{-1}$ &   103 & $\sim 10^{-2}$ &   104 \\
$7d_{3/2}$ & -1350 & $\sim 10^{-3}$ & -106 & $\sim 10^{-3}$ &   284 & $\sim 10^{-3}$ &   289 \\
$7d_{5/2}$ &  -606  & $\sim 10^{-8}$ &  -48 & $\sim 10^{-7}$ &   128 & $\sim 10^{-8}$ &   130 \\
\end{tabular}
\end{ruledtabular}
\end{table*}

A hole (or, more accurately, central depression) in nuclear density for E120 was considered in Refs.~\cite{Bender,Decharge,Afanasiev1}. Its importance is related to theoretical prediction of magic numbers for protons and neutrons. 
We study the effect of making a hole in nuclear charge distribution by comparing the energies of the $^{292}$E120$^+$ ion in which nuclear charge distribution is pure Fermi distribution (\ref{e:Fermi}) to the energies of the ion in which nuclear density is modified according to (\ref{e:hole}) (see also Fig.~\ref{f:dro}). We use $k=0.5$ while keeping the RMS radius fixed. The results are presented in Table~\ref{t:IS}. We also present in this table reference IS which is the shift between $^{292}$E120 and $^{294}$E120 calculated with the nuclear parameters from Table~\ref{t:E120} as a matrix element $\langle \psi_a^{\rm Br}| \delta V_N + \delta V_{core} | \psi_a^{\rm Br} \rangle$. The ratio of the energy shifts due to a hole to the reference IS is about 8\%. This means that the effect is significant and deserves further study. 

It turns out that a hole in the nuclear charge distribution  is numerically equivalent to decreasing the value of the 
skin thickness (parameter $t$ in (\ref{e:Fermi})). The value $k=0.5$ corresponds to the 14.5\% decrease in 
the value of $t$. In both cases the effect is practically the same for all considered states. 

\subsection{Nuclear quadrupole deformation and change of nuclear radius}

Next we study the effect of nuclear quadrupole deformation. We consider a model situation by comparing two nuclei with the same RMS radius but one  has no deformation, and another has a deformation with $\beta=-0.4$ in (\ref{e:beta}). This value of $\beta$ comes from nuclear calculations for the $^{316}$E120 isotope~\cite{Afanasiev}. The effect of
quadrupole deformation is equivalent to increased skin thickness (see Fig.~\ref{f:dro}). Calculations show that
for $\beta=-0.4$ equivalent increase in skin thickness$t$  is 30.5\% in good agreement with (\ref{e:beta1}). 
The shift in energy is significant, $\sim 2~{\rm cm}^{-1}$ for $s$ states (see Table~\ref{t:IS}) or $\sim$~20\% of the reference IS for all considered states. 
This leads to a question whether IS can be used to study nuclear deformation. Therefore, we check whether nuclear deformation can be
distinguished  from the change of nuclear RMS radius. Two last columns of Table~\ref{t:IS} show the effect of the change 
in nuclear RMS radius in which the parameters were chosen to produce the same IS for the $8s$ state as in the case of 
quadrupole deformation. We see that the shift is the same for all states except the $8p_{1/2}$ state. The difference 
for the $8p_{1/2}$ state is 4\% or 0.014~cm$^{-1}$. This is large enough to be detected in spectroscopic measurements.
However, this is a model case. Let us now consider a more realistic case of isotope shift between two isotopes in which nuclear parameters are taken from nuclear calculations~\cite{Afanasiev}. We consider isotope shift for frequencies of electric dipole transitions in 
E120$^+$ for isotopes in Table~\ref{t:E120}. IS for the $a \rightarrow b$ transition is given by $\delta \nu_{ab} = \langle \psi_b^{\rm Br}| \delta V_N + \delta_{core}| \psi_b^{\rm Br} \rangle  - \langle \psi_a^{\rm Br}| \delta V_N +\delta V_{core}| \psi_a^{\rm Br} \rangle$.
The results are presented as case A in Table~\ref{t:w}. In case B we
perform model calculations to check whether IS can be reduced to the change in RMS radius. The answer is negative. 
We see that if we chose the change in RMS radius to fit the shift of $s$ and $p_{3/2}$ states (they behave the same way, see
above) then the shift for the $p_{1/2}$ state is slightly different leading to different IS in the $ns-mp_{1/2}$ transitions.
The difference is $\sim 0.003~{\rm cm}^{-1}$ for the $8s - 8p_{1/2}$ transition which is probably large enough to be 
detected. This means that nuclear deformation can be studied by comparing IS in $s - p_{1/2}$ and $s - p_{3/2}$
transitions. Both these IS cannot be fitted by changing just one nuclear parameter, e.g. RMS radius. Change in nuclear
deformation ($\beta$) is also needed. Note that this might be the only way of  study  nuclear 
deformation for even-even isotopes by means of atomic spectroscopy. In odd isotopes one can also measure electric quadrupole hyperfine structure.
Note also that since three types of nuclear deformations (hole in the origin, quadrupole deformation and change of thickness) are numerically equivalent in terms of producing similar IS, what is said above about nuclear deformation is also true about having a hole in nuclear charge distribution;
i.e. it can be studied by comparing IS in $s - p_{1/2}$ and $s - p_{3/2}$ transitions.

\begin{table}
\caption{\label{t:w}Isotope shift (in cm$^{-1}$) for the frequencies of the $8s-8p$ and $9s-8p$ transitions in E120$^+$. 
Case A corresponds to nuclear parameters in Table~\ref{t:E120}. Case B is a model case in which $\beta=0$ 
for both isotopes and change in RMS radius is chosen to fit the shift of $s$ states.}
\begin{ruledtabular}
\begin{tabular}{lddd}
\multicolumn{1}{c}{Transition}&
\multicolumn{1}{c}{A}&
\multicolumn{1}{c}{B}&
\multicolumn{1}{c}{A-B}\\
\hline
$8s - 8p_{1/2}$ &  8.42911  & 8.43187  & -0.0028 \\
$8s - 8p_{3/2}$ & 10.6185  & 10.6183   & 0.0002 \\

$9s - 8p_{1/2}$ &  0.67423 & 0.67528  & -0.0011 \\
$9s - 8p_{3/2}$ &  2.86118  & 2.86109  & 0.0001 \\

\end{tabular}
\end{ruledtabular}
\end{table}

\subsection{Isotope shift for large change of neutron numbers}

\begin{figure}
\epsfig{figure=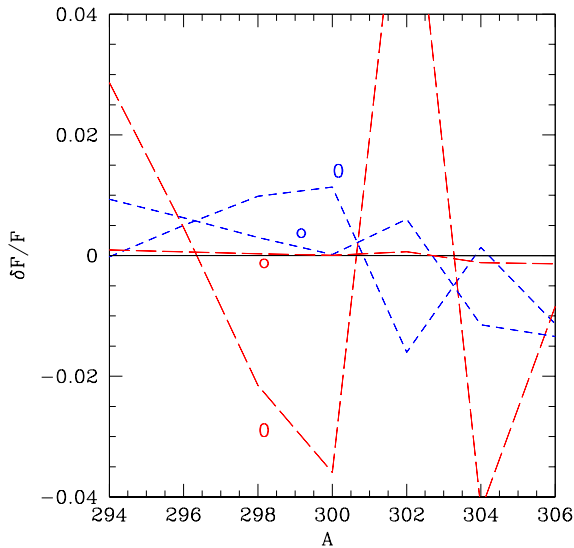,scale=0.4}
\caption{Fractional deviation from average value for isotope shift ratio (black solid line); field shift constant $F$ (blue short dashed lines) and modified field shift constant $\tilde F$ (red long dashed lines) for spherically symmetric and deformed nuclei. Lines, corresponding to spherically symmetric nuclei market with "o"; lines corresponding to deformed nuclei market with "0".
}
\label{f:dv0}
\end{figure}

It was suggested in Ref.~\cite{dzuba_isotope_2017} to use isotope shift calculations to predict transition frequencies in SHE from a hypothetical island of stability. These metastable SHE differ from isotope-poor SHE produced in laboratories by large number of neutrons (large $\Delta N$). This should be taken into account in the IS calculations. Calculations reported above use the RPA method which assumes that the change in nuclear potential $\delta V_N$ is a small perturbation and ignores non-linear in $\delta V_N$ contributions. In SHE with large $\Delta N$ non-linear in $\delta V_N$ contributions are likely to be important and should not be thrown away. The most obvious way to do calculations properly  is to calculate energy levels for each isotope and then take the difference. This does not work for light atoms because the IS is small and obtaining it as a difference of large almost equal numbers leads to numerical instabilities. Fortunately, IS in SHE is sufficiently large to ensure stable results. Even for neighbouring isotopes taking the difference between two RHF calculations produce  result which are very close to the RPA calculations. For large $\Delta N$, the calculations based on the difference between two isotopes are preferable because they include non-linear contributions.

It is customary to present FIS as a formula in which electron and nuclear variables are separated.
Standard formula reads
\begin{equation}\label{e:fis}
{\rm FIS}= F \delta\langle r^2 \rangle.
\end{equation}
It is assumed that the electron structure factor $F$ does not depend on nuclear variables. This formula works very well in light atoms and widely used even for atoms close to the end of known periodic table (e.g. for No, $Z=102$,~\cite{No}).
It was shown in Ref.~\cite{FlambaumGeddesViatkina} that relativistic corrections lead to a different formula
\begin{equation}\label{e:fism}
{\rm FIS}= \tilde F \delta\langle r^{2\gamma} \rangle,
\end{equation}
where $\gamma = \sqrt{1-(\alpha Z)^2}$. New electron structure constant $\tilde F$ does not depend on nuclei. The formula was obtained by considering spherical nuclei with uniform change distribution. Below we study the performance of both formulae  (\ref{e:fis}) and (\ref{e:fism}) for deformed nuclei.
We calculate  isotope shifts for the $8s -8p_{1/2}$ and $8s -8p_{3/2}$ transitions for all even isotopes of E120$^+$ from $A$=294 to $A$=306.
We take nuclear parameters $\beta$ and RMS radius from Ref.~\cite{Afanasiev} (see Table~\ref{t:E120}). We also consider a model case in which all considered
nuclei are assumed to be spherically symmetric ($\beta=0$). 
IS is calculated for pairs of neighbouring isotopes using the RPA method as described above. The constants $F$ and $\tilde F$ are found using (\ref{e:fis}) and (\ref{e:fism}). The calculations repeated for both transitions for seven pairs of isotopes. In the end we have fourteen values of isotope shift and fourteen values of $F$ and $\tilde F$.
The results are presented on Fig.~\ref{f:dv0} in terms of fractional deviations of the considered values from their average values, e.g. $\delta (F/F)_i= (F_i-\langle F \rangle)/\langle F \rangle$, where $\langle F \rangle = \sum F_i/7$.  
We present on Fig.~\ref{f:dv0} the variation of the ratio of isotope shifts in two transitions and variations of $F$ and $\tilde F$ for both transitions. 
However, the variations for two transitions are too similar to see the difference on the graph.
Fig.~\ref{f:dv0} shows that the ratio of the isotope shifts remains constant to very high precision. 
However, neither formula (\ref{e:fis}) or (\ref{e:fism}) works well. The value of $F$ in (\ref{e:fis}) tends to drift in one direction leading to large variations for large difference in neutron numbers. This is similar for both cases, symmetrical and deformed nuclei. In contrast, formula (\ref{e:fism}) works very well for spherical nuclei, showing only about 0.01\% variation for $\tilde F$ in the considered interval. However, the formula does not work so well for deformed nuclei. The value for $\tilde F$ jumps up and down by several per cent from one isotope to another. This is probably because the value of $\langle r^{2\gamma} \rangle$ depends on two nuclear parameters, nuclear deformation parameter $\beta$ and nuclear RMS radius, making its behaviour irregular.

Note that the difference in the value of $F$ for neighbouring isotopes usually does not exceed 1\% for both spherical and deformed nuclei. With this accuracy formula (\ref{e:fis}) can be used for neighbouring isotopes to extract the change of nuclear RMS radius from isotope shift measurements (see, e.g.~\cite{No}). Keeping in mind that the value of $F$ depends on isotope, the calculations should be performed for one of isotopes of interest (or for both, taking then an average value).

\section{Conclusions}

We studied the effects of nuclear deformations on the field isotope shift in SHE. We demonstrated that making a hole in nuclear charge distribution and having quadrupole deformation can be reduced to changing nuclear skin thickness. On the other hand, changing in skin thickness is not totally equivalent to change of nuclear RMS radius. There is small difference in energy shift of the $p_{1/2}$ states compared to states of other symmetries.
With sufficiently accurate measurements of the IS this difference can probably be used to study nuclear deformations in even nuclei.
The total effect of the nuclear hole on the isotope shift is up to  $\sim 8 \%$, the effect of the deformation is up to $\sim 20\%$.
  
  We demonstrated that known formulae for separation of nuclear and electron variables do not work for heavy deformed nuclei. However, in considered examples the ratio of isotope shifts for two atomic transitions remained isotope-independent. 
Therefore, the  linearity of King plot is not broken.

\section{Acknowledgements}
This work was supported by the Australian Research Council and the Gutenberg Fellowship.
The authors are grateful to Anatoli Afanasjev, Jie Meng, Baohua Sun, Jorge Piekarewicz,  Shashi K. Dhiman, Jacek Dobaczewski, Peter Ring, Stephane Goriely, Zhongzhou Ren, Ning Wang, Yibin Qian, Witold Nazarewicz, Paul-Gerhard Reinhard, Peter Schwerdtfeger  and Bryce Lackenby  for valuable discussions.

\bibliographystyle{apsrev}

\begin{thebibliography}{47}
\expandafter\ifx\csname natexlab\endcsname\relax\def\natexlab#1{#1}\fi
\expandafter\ifx\csname bibnamefont\endcsname\relax
  \def\bibnamefont#1{#1}\fi
\expandafter\ifx\csname bibfnamefont\endcsname\relax
  \def\bibfnamefont#1{#1}\fi
\expandafter\ifx\csname citenamefont\endcsname\relax
  \def\citenamefont#1{#1}\fi
\expandafter\ifx\csname url\endcsname\relax
  \def\url#1{\texttt{#1}}\fi
\expandafter\ifx\csname urlprefix\endcsname\relax\def\urlprefix{URL }\fi
\providecommand{\bibinfo}[2]{#2}
\providecommand{\eprint}[2][]{\url{#2}}

\bibitem[{\citenamefont{Oganessian et~al.}(2004)\citenamefont{Oganessian,
  Utyonkov, Lobanov, Abdullin, Polyakov, Shirokovsky, Tsyganov, Gulbekian,
  Bogomolov, Gikal et~al.}}]{oganessian_heavy_2004}
\bibinfo{author}{\bibfnamefont{Y.~T.} \bibnamefont{Oganessian}},
  \bibinfo{author}{\bibfnamefont{V.~K.} \bibnamefont{Utyonkov}},
  \bibinfo{author}{\bibfnamefont{Y.~V.} \bibnamefont{Lobanov}},
  \bibinfo{author}{\bibfnamefont{F.~S.} \bibnamefont{Abdullin}},
  \bibinfo{author}{\bibfnamefont{A.~N.} \bibnamefont{Polyakov}},
  \bibinfo{author}{\bibfnamefont{I.~V.} \bibnamefont{Shirokovsky}},
  \bibinfo{author}{\bibfnamefont{Y.~S.} \bibnamefont{Tsyganov}},
  \bibinfo{author}{\bibfnamefont{G.~G.} \bibnamefont{Gulbekian}},
  \bibinfo{author}{\bibfnamefont{S.~L.} \bibnamefont{Bogomolov}},
  \bibinfo{author}{\bibfnamefont{B.~N.} \bibnamefont{Gikal}},
  \bibnamefont{et~al.}, \bibinfo{journal}{Nuclear Physics A}
  \textbf{\bibinfo{volume}{734}}, \bibinfo{pages}{109} (\bibinfo{year}{2004}).

\bibitem[{\citenamefont{Hamilton et~al.}(2013)\citenamefont{Hamilton, Hofmann,
  and Oganessian}}]{hamilton_search_2013}
\bibinfo{author}{\bibfnamefont{J.~H.} \bibnamefont{Hamilton}},
  \bibinfo{author}{\bibfnamefont{S.}~\bibnamefont{Hofmann}}, \bibnamefont{and}
  \bibinfo{author}{\bibfnamefont{Y.~T.} \bibnamefont{Oganessian}},
  \bibinfo{journal}{Annual Review of Nuclear and Particle Science}
  \textbf{\bibinfo{volume}{63}}, \bibinfo{pages}{383} (\bibinfo{year}{2013}).

\bibitem[{\citenamefont{Fuller et~al.}(2017)\citenamefont{Fuller, Kusenko, and
  Takhistov}}]{fuller_primordial_2017}
\bibinfo{author}{\bibfnamefont{G.~M.} \bibnamefont{Fuller}},
  \bibinfo{author}{\bibfnamefont{A.}~\bibnamefont{Kusenko}}, \bibnamefont{and}
  \bibinfo{author}{\bibfnamefont{V.}~\bibnamefont{Takhistov}},
  \bibinfo{journal}{Physical Review Letters} \textbf{\bibinfo{volume}{119}},
  \bibinfo{pages}{061101} (\bibinfo{year}{2017}).

\bibitem[{\citenamefont{Goriely et~al.}(2011)\citenamefont{Goriely, Bauswein,
  and Janka}}]{goriely_r-process_2011}
\bibinfo{author}{\bibfnamefont{S.}~\bibnamefont{Goriely}},
  \bibinfo{author}{\bibfnamefont{A.}~\bibnamefont{Bauswein}}, \bibnamefont{and}
  \bibinfo{author}{\bibfnamefont{H.-T.} \bibnamefont{Janka}},
  \bibinfo{journal}{The Astrophysical Journal Letters}
  \textbf{\bibinfo{volume}{738}}, \bibinfo{pages}{L32} (\bibinfo{year}{2011}).

\bibitem[{\citenamefont{Frebel and Beers}(2018)}]{frebel_formation_2018}
\bibinfo{author}{\bibfnamefont{A.}~\bibnamefont{Frebel}} \bibnamefont{and}
  \bibinfo{author}{\bibfnamefont{T.~C.} \bibnamefont{Beers}},
  \bibinfo{journal}{Physics Today} \textbf{\bibinfo{volume}{71}},
  \bibinfo{pages}{30} (\bibinfo{year}{2018}), ISSN \bibinfo{issn}{0031-9228}.

\bibitem[{\citenamefont{Gopka et~al.}(2008)\citenamefont{Gopka, Yushchenko,
  Yushchenko, Panov, and Kim}}]{gopka_identification_2008}
\bibinfo{author}{\bibfnamefont{V.~F.} \bibnamefont{Gopka}},
  \bibinfo{author}{\bibfnamefont{A.~V.} \bibnamefont{Yushchenko}},
  \bibinfo{author}{\bibfnamefont{V.~A.} \bibnamefont{Yushchenko}},
  \bibinfo{author}{\bibfnamefont{I.~V.} \bibnamefont{Panov}}, \bibnamefont{and}
  \bibinfo{author}{\bibfnamefont{C.}~\bibnamefont{Kim}},
  \bibinfo{journal}{Kinematics and Physics of Celestial Bodies}
  \textbf{\bibinfo{volume}{24}}, \bibinfo{pages}{89} (\bibinfo{year}{2008}).

\bibitem[{\citenamefont{Dzuba et~al.}(2017)\citenamefont{Dzuba, Flambaum, and
  Webb}}]{dzuba_isotope_2017}
\bibinfo{author}{\bibfnamefont{V.~A.} \bibnamefont{Dzuba}},
  \bibinfo{author}{\bibfnamefont{V.~V.} \bibnamefont{Flambaum}},
  \bibnamefont{and} \bibinfo{author}{\bibfnamefont{J.~K.} \bibnamefont{Webb}},
  \bibinfo{journal}{Physical Review A} \textbf{\bibinfo{volume}{95}},
  \bibinfo{pages}{062515} (\bibinfo{year}{2017}).

\bibitem[{\citenamefont{Witten}(1984)}]{witten_cosmic_1984}
\bibinfo{author}{\bibfnamefont{E.}~\bibnamefont{Witten}},
  \bibinfo{journal}{Physical Review D} \textbf{\bibinfo{volume}{30}},
  \bibinfo{pages}{272} (\bibinfo{year}{1984}).

\bibitem[{\citenamefont{Berengut et~al.}(2017)\citenamefont{Berengut, Budker,
  Delaunay, Flambaum, Frugiuele, Fuchs, Grojean, Harnik, Ozeri, Perez
  et~al.}}]{berengut_probing_2017}
\bibinfo{author}{\bibfnamefont{J.~C.} \bibnamefont{Berengut}},
  \bibinfo{author}{\bibfnamefont{D.}~\bibnamefont{Budker}},
  \bibinfo{author}{\bibfnamefont{C.}~\bibnamefont{Delaunay}},
  \bibinfo{author}{\bibfnamefont{V.~V.} \bibnamefont{Flambaum}},
  \bibinfo{author}{\bibfnamefont{C.}~\bibnamefont{Frugiuele}},
  \bibinfo{author}{\bibfnamefont{E.}~\bibnamefont{Fuchs}},
  \bibinfo{author}{\bibfnamefont{C.}~\bibnamefont{Grojean}},
  \bibinfo{author}{\bibfnamefont{R.}~\bibnamefont{Harnik}},
  \bibinfo{author}{\bibfnamefont{R.}~\bibnamefont{Ozeri}},
  \bibinfo{author}{\bibfnamefont{G.}~\bibnamefont{Perez}},
  \bibnamefont{et~al.}, \bibinfo{journal}{arXiv:1704.05068}  (\bibinfo{year}{2017}),
  \bibinfo{note}{(accepted by Phys. Rev. Lett.)}.

\bibitem{FlambaumGeddesViatkina} Isotope shift, nonlinearity of King plots, and the search for new particles
V. V. Flambaum, A. J. Geddes, and A. V. Viatkina
Phys. Rev. A 97, 032510 (2018).

\bibitem{No} S. Raeder, D. Ackermann, H. Backe, {\em et al}, Phys. Rev. Lett. {\bf 120}, 232503 (2018).

\bibitem{Db} B. G. C. Lackenby, V. A. Dzuba, and V. V. Flambaum,
Phys. Rev. A {\bf 98}, 022518 (2018).

\bibitem{Sg-Mt} B. G. C. Lackenby, V. A. Dzuba, and V. V. Flambaum,
Phys. Rev. A {\bf 99}, 042509 (2019). 


\bibitem[{\citenamefont{Racah}(1932)}]{racah_isotopic_1932}
\bibinfo{author}{\bibfnamefont{G.}~\bibnamefont{Racah}},
  \bibinfo{journal}{Nature} \textbf{\bibinfo{volume}{129}},
  \bibinfo{pages}{723} (\bibinfo{year}{1932}).

\bibitem[{\citenamefont{Rosenthal and Breit}(1932)}]{rosenthal_isotope_1932}
\bibinfo{author}{\bibfnamefont{J.~E.} \bibnamefont{Rosenthal}}
  \bibnamefont{and} \bibinfo{author}{\bibfnamefont{G.}~\bibnamefont{Breit}},
  \bibinfo{journal}{Physical Review} \textbf{\bibinfo{volume}{41}},
  \bibinfo{pages}{459} (\bibinfo{year}{1932}).

\bibitem[{\citenamefont{King}(2013)}]{king_isotope_2013}
\bibinfo{author}{\bibfnamefont{W.~H.} \bibnamefont{King}},
  \emph{\bibinfo{title}{Isotope {Shifts} in {Atomic} {Spectra}}}
  (\bibinfo{publisher}{Springer Science \& Business Media},
  \bibinfo{year}{2013}).

\bibitem[{\citenamefont{Shabaev}(1993)}]{shabaev_finite_1993}
\bibinfo{author}{\bibfnamefont{V.~M.} \bibnamefont{Shabaev}},
  \bibinfo{journal}{Journal of Physics B: Atomic, Molecular and Optical
  Physics} \textbf{\bibinfo{volume}{26}}, \bibinfo{pages}{1103}
  (\bibinfo{year}{1993}).
 

\bibitem{Afanasiev} S. E. Agbemava, A. V. Afanasjev, T. Nakatsukasa, and P. Ring,
Phys. Rev. C {\bf 92}, 054310 (2015). 

\bibitem{E120}  T. H. Dinh, V. A. Dzuba, V. V. Flambaum and J. S. M. Ginges,
       Phys. Rev. A {\bf 78}, 022507 (2008).
\bibitem{E120a} V. A. Dzuba,
      Phys. Rev. A {\bf 88}, 042516 (2013).  
\bibitem{Sigma} V. A. Dzuba, V. V. Flambaum, P. G. Silvestrov, O. P. Sushkov,
J. Phys. B: {\it At. Mol. Phys.}, 
{\bf 20}, 1399-1412 (1987).

\bibitem{Bender} M. Bender, K. Rutz, P.-G. Reinhard, J. A. Maruhn, and W. Greiner, Phys. Rev. C {\bf 60}, 034304 (1999).

\bibitem{Decharge} J. Decharge, J.-F. Berger, K. Dietrich, and M. S. Weiss, Phys. Lett. B {\bf 451}, 275 (1999).

\bibitem{Afanasiev1} A. V. Afanasjev and S. Frauendorf,
Phys. Rev. C {\bf 71}, 024308 (2005). 

\bibitem{Clark} D. L. Clark, M. E. Cage, D. A. Lewis, and G. W. Greenlees, Phys. Rev. A {\bf 20}, 239 (1979).

\bibitem{Heisenberg} J. H. Heisenberg, J. S. McCarthy, I. Sick, and M. R. Yearian, Nucl.  Phys. A {\bf 164}, 340 (1971).

\end{thebibliography}

\end{document}